\documentclass[aps,prl,twocolumn,showpacs,superscriptaddress,groupedaddress]{revtex4-1}  
\usepackage{graphicx}  
\usepackage{dcolumn}   
\usepackage{bm}        
\usepackage{amssymb}   
\usepackage{subfigure}
\usepackage{amsmath}
\usepackage{float}
\usepackage{blindtext}
\usepackage[toc,page]{appendix}

\begin{document}

\title{Discontinuous shear thickening of dense suspensions under confining pressure}

\author{Junhao Dong}
\author{Martin Trulsson}
\affiliation{Theoretical Chemistry, Department of Chemistry, Lund University, Sweden}

\date{\today}

\begin{abstract}
We use 2D numerical simulations to study dense suspensions of non-Brownian hard particles using the Critical Load Model (CLM) under constant confining pressures.
This simple model shows discontinuous shear thickening (DST) as the tangential forces get activated upon increased shear stresses. 
By parameterizing a simple binary system of frictional and non-frictional particles of different proportions we show that the jamming packing fraction, at which the viscosity diverges, is controlled by the fraction of frictional contacts.
The viscosity of dense suspensions can thereby be expressed as a function of the fraction of frictional contacts as well as the packing fraction of solid particles. 
In addition, we show that there exists a simple relationship between the fraction of frictional contacts and the two control parameters (under confining pressure): the viscous number $J$ and the ratio between the repulsive barrier force and confining pressure.
Under confining pressures the viscosity curves are found to depend on the shear protocol, with the possibility of yielding negative \textit{dynamic} compressibility.
\end{abstract}


\maketitle

Dense suspensions of rigid particles are of great importance both from a geological and an industrial point of view, including materials such as mud, quicksand, slurry, tooth-paste, paint, etc. Such suspensions a variety of rheological behaviors and a substantial amount of research have been done trying to facilitate our understanding of these behaviors \cite{Mari3}. 

In the simplest model describing non-Brownian suspensions, where particles are assumed to have no inertia or Brownian motion and being immersed in a highly viscous fluid, the suspension's viscosity will be a sole function of
packing fraction $\phi$ and the fluid viscosity \cite{Boyer1}. While the viscosity of dilute suspensions is given by the hydrodynamic stresses on single particles in a shear flow and distant hydrodynamic interactions between pairs \cite{Batchelor1} 
the viscosity in dense suspensions is governed by viscous dissipation due to particle departures from the affine flow due to geometrical constraints \cite{Andreotti1,Lerner1,Trulsson2}. The path of departures will increase as the packing fraction increases and finally diverge
at the jamming transition. While this is a good model for most of suspensions, it does neither include the possibility of shear thinning nor shear thickening. The latter phenomenon has lately acquired an increasing interest in basic research. In these suspensions the viscosity increases as a function of shear rate \cite{Barnes1}. This increase is labelled as continuous or discontinuous depending on how significant and how sharp the increase is. Transition with significant and sharp viscosity increase is defined as discontinuous shear thickening (DST). DST is usually observed when packing fraction of the suspension exceeds a certain threshold while lower packing fraction gives a more continuous shear thickening \cite{Seto1}. Several scenarios have been proposed to explain shear thickening behavior.  It is known that the inertia of the particles \cite{Bagnold1} and/or fluid \cite{Picano1} gives a continuous shear thickening above certain shear rate, usually described by the Stokes and/or Reynolds number(s). However, other scenarios are also proposed to explain shear thickening with inertia being subdominant. 

Hydroclustering has been a dominating explanation for a couple of decades, where shear thickening is driven by clustering of particles upon increasing shear rate leading to effectively larger quasi-particles.  These transient hydroclusters have little internal re-arrangement but a large collective rotational and translational movement which leads to a high dissipation, hence increased viscosity \cite{Brady1,Farr1}.  Although hydro-clustering works well in describing weak shear thickening, simulations based on this phenomenon have not been able to produce the dramatic viscosity increase observed in discontinuous shear thickening \cite{Brown1}. 
Another alternative explanation relies on the onset of tangential force with increasing shear rate or, equivalently increasing stresses \cite{Fernandez1, Seto1}.  It is now well-studied that the rheology of frictional and non-frictional dense suspension differ in the jamming packing fraction and the viscosity at the same shear-rate, where frictional suspensions diverge at lower packing fractions and have higher viscosities at the same packing fraction compared to non-frictional suspensions \cite{Mari3}. 
The former results from a counting argument where extra frictional forces can mechanically stabilise the packings with fewer contacts per particle, hence lower packing fractions.
These extra tangential forces also increase the dissipation and hence the viscosity.

In the frictional explanation of DST one goes from a non-frictional to a frictional suspension as the frictional contacts are "activated".  This activation is initiated once the particles have enough energy to overcome a certain repulsive barrier that protect the particles from being in contact and feeling the frictional forces. Such repulsive and lubricative forces could be both electrostatic and/or steric, \textit{e.g.} polymer brushes. Several recent papers have reported simulation results based on this explanation/model which produce DST behavior. 
Although there has been a lot of attention following this explanation/model, most works are done under constant packing fraction \cite{Wyart1, Seto1, Mari2, Ness1} and very little is known about how these suspensions behave under constant confining pressure, which might be a more relevant boundary condition for geological processes or non-planar shear cells \cite{Fall1}.

In our work, we explore behavior of dense suspensions under confining pressure based on the scenario of "activation" of tangential force. We focus on non-Brownian suspensions in highly viscous fluids. Besides of finding relations between viscosity, packing fraction and fraction of frictional contact, we also explore different shear protocols yielding different viscosity versus packing fraction curves. 

Simulations are run in 2D with a constant number of disks, $N_\mathrm{p}=948$, with average diameter $d$ and polydispersity $\pm50\%$ to avoid crystallisation. Periodic boundary condition is applied along the $x-$direction. The cell size is approximately $47d$ in the $x$-direction (along the shear) but is allowed to vary in the other direction (typically $\sim 20d$). The disks are confined between two rough walls and immersed in a highly viscous fluid with viscosity $\eta_0$. The disk dynamics is hence strictly overdamped without Brownian noise which gives us a set of equations of force and torque balance for each disk $i$. The force balance is given as $\vec F^\mathrm{ext}_i+\vec F^\mathrm{v}_i-\sum_j{ \vec f_{ij}}=0$, where $F^\mathrm{ext}$ is the external force from walls, $\vec F^\mathrm{v}$ is the viscous drag force, and $f_{ij}$ is the contact force exerted on disk $j$ by disk $i$. The drag force is given by Stokes drag $\vec F^\mathrm{v}_i=-\frac{3\pi\eta_0}{1-\phi_0}(\vec V_i-\vec V_i^\mathrm{a})$, where $\vec V_i^\mathrm{a} =\dot \gamma y \hat x$ is the affine fluid velocity at a shear-rate $\dot \gamma$ and $\phi_0=0.76$. The torque balance is $\tau^\mathrm{ext}_i+\tau^\mathrm{v}_i-\sum_j{\tau_{ij}}=0$. The force between two disks consist of two components: normal force $\vec f_\mathrm{n}$ and tangential force $\vec f_\mathrm{t}$. Here, we assume elastic stiff disks in our simulation so the normal force is given by a linear force, $f_\mathrm{n}^{ij}=k_\mathrm{n}\delta_\mathrm{n}^{ij}$, where $\delta_\mathrm{n}^{ij}$ is the overlap between two disks and $k_\mathrm{n}$ is the normal spring constant. The tangential force is obtained in a similar way but using tangential spring and  a tangential spring constant $k_\mathrm{t}=0.5k_\mathrm{n}$. The relation between normal component and tangential component is constrained by Coulomb friction, $|{f_\mathrm{t}}|\leq\mu_\mathrm{p}|{f_\mathrm{n}}|$, where $\mu_p$ is the friction coefficient (for more information see \textit{e.g.} \cite{Trulsson1}).

We use the Critical Load Model (CLM) \cite{Mari3, Seto1, Ness1} which generates DST behavior. This model describes transition between lubrication (non-frictional) and frictional contacts. We first define a critical normal force $f_\mathrm{n}^\mathrm{cl}$ which represents the magnitude of the repulsive force between two disks at which frictional contacts set in. The friction coefficient between disk $i$ and $j$ is then determined by comparing the normal force between two disks with $f_\mathrm{n}^\mathrm{cl}$,
\begin{equation}\label{eq:mup}
\mu_\mathrm{p}^{ij}=
\begin{cases}
1,& f_\mathrm{n}^{ij}>f_\mathrm{n}^\mathrm{cl};\\ 0,& f_\mathrm{n}^{ij}<f_\mathrm{n}^\mathrm{cl}.
\end{cases}
\end{equation}
Accordingly, a contact is defined as frictional when $\mu_\mathrm{p}^{ij}=1$; otherwise, it is frictionless.  We define the fraction of frictional contact $\chi_\mathrm{f}$ as the ratio between number of frictional contacts and the number of total contact. In order to study the influence of $\chi_\mathrm{f}$ on viscosity of the suspension,  we also run simulation with constant $\chi_\mathrm{f}$, where disks have a binary distribution of friction coefficient. On this condition, we set certain fraction of disks, $N_\mathrm{f}$,  as frictional, \textit{i.e.}~$\mu_\mathrm{p}^i=1$, while the others are frictionless, \textit{i.e.}~$\mu_\mathrm{p}^i=0$.  The friction coefficient between disk $i$ and $j$ is defined as $\mu_\mathrm{p}^{ij}=\sqrt{\mu_\mathrm{p}^i\mu_\mathrm{p}^j}$ (\textit{i.e.} only between two frictional disks can one have frictional contacts). In this way, it is easy to control the fraction of frictional contacts $\chi_\mathrm{f}\simeq{N_\mathrm{f}^2}$. 

The simulations are shear rate-controlled, by imposing constant velocities of the walls, either at constant packing fraction or constant pressure. The latter is fulfilled by imposing a constant pressure $P$ to both walls, allowing the shear cell to dilate or
compress during the shearing. The disk velocities follow well the linear profile, with slope $\dot \gamma$, set by the viscous fluid. All the measurements are done considering only the central part of the shear cell, excluding five layers of disks close to each wall. In our simulation, we vary the viscous number $J=\eta_0\dot{\gamma}/P$ by changing either the confining pressure $P$ or the shear rate $\dot{\gamma}$ and measure how the viscosity $\eta=\sigma/\dot{\gamma}$, where $\sigma$ is shear stress, and $\chi_\mathrm{f}$ changes with $J$. The stiffness of disks is maintained by keeping $k_\mathrm{n}/P\approx2\cdot10^3$ (giving in principal a hard disk behavior). To get more systematic understanding, we run simulations with either $f_\mathrm{n}^\mathrm{cl}/(Pd)$ or $f_\mathrm{n}^\mathrm{cl}/(\eta_0\dot \gamma d)$ constant corresponding to two different shear protocols; keeping either $P$ constant and varying $\dot \gamma$ or keeping $\dot \gamma$ constant and varying $P$. We also run two reference simulations; in one case, all contacts are frictional (\textit{i.e.}~$f_\mathrm{n}^\mathrm{cl}=0$) while in the other case, all contacts are frictionless (\textit{i.e.}~$f_\mathrm{n}^\mathrm{cl}=\infty$). 


To validate that the model used is able to describe DST behavior, we first run the simulation with four constant packing fraction $\phi=0.43, 0.54, 0.63, 0.76$. We plot viscosity as a function of shear rate in normalised units, $\eta/\eta_0$ against $\dot{\gamma}\eta_0/(f_\mathrm{n}^\mathrm{cl}d)$ (see Supplementary Information \cite{SI}). All the curves clearly show a discontinuous transition in viscosity when the shear rate reaches a critical value $\dot{\gamma_0}$. At $\phi=0.43$, the increase is quite small. As $\phi$ increases, the transition becomes more pronounced. This result is consistent with previous reported works \cite{Seto1, Wyart1}.

\begin{figure}[t]
\includegraphics[scale=0.4]{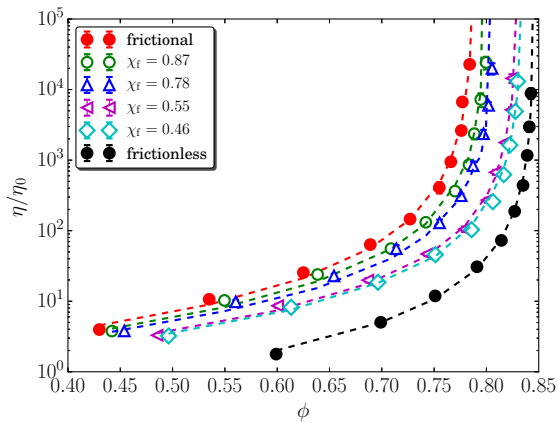}
\caption{\label{fig:eta_phi_frac}  Viscosity $\eta/\eta_0$ as a function of packing fraction $\phi$ for various $\chi_\mathrm{f}$. Symbols are simulation results and dashed lines are plots of Eq.~\ref{eq:eta-chi} with corresponding $\chi_\mathrm{f}$.}
\end{figure}

We then run simulations with a constant confining pressure $P$ and study how the viscosity $\eta$ changes with packing fraction $\phi$. Divergence of dense suspension close to jamming can be estimated as
\begin{equation}\label{eq:eta-phi}
\eta/\eta_0\simeq{k(\phi_\mathrm{c}-\phi)^{-m}}, 
\end{equation}
where $\phi_\mathrm{c}$ is critical packing fraction where jamming transition happens. For a constant $\chi_\mathrm{f}$, $\phi$ is a function of $J$  \cite{Boyer1}; $\phi\to\phi_\mathrm{c}$ when $J\to0$. 
Since frictional contact will influence packing of disks, it is reasonable to think that $\phi_\mathrm{c}=\phi_\mathrm{c}(\chi_\mathrm{f})$ and bound by the two limits $\phi_\mathrm{c}(0)=\phi_\mathrm{c}^\mathrm{nf}$ and $\phi_\mathrm{c}(1)=\phi_\mathrm{c}^\mathrm{f}$, where the subscripts denote either the fully frictional or non-frictional references.
All intermediate $\chi_\mathrm{f}$ cases can then be expressed as
\begin{equation}\label{eq:phic}
\phi_\mathrm{c}(\chi_\mathrm{f})=\phi_\mathrm{c}^\mathrm{f}g(\chi_\mathrm{f})+\phi_\mathrm{c}^\mathrm{nf}\big(1-g(\chi_\mathrm{f})\big),
\end{equation}
 where $g(\chi_f)\in[0,1]$ is a combination function conjectured by Wyart and Cates \cite{Wyart1}. Furthermore, we assume that  $k=k(\chi_\mathrm{f})$ and $m=m(\chi_\mathrm{f})$ so that
\begin{equation}\label{eq:eta-chi}
\eta/\eta_0=\eta(\phi, \chi_\mathrm{f})=k(\chi_\mathrm{f}) \big(\phi_\mathrm{c}(\chi_\mathrm{f})-\phi\big)^{-m(\chi_\mathrm{f})}.
\end{equation}
To test our assumption, we first run simulations with constant $\chi_\mathrm{f}$ (binary system). The results are plotted in Fig.~\ref{fig:eta_phi_frac}. We fit the data using Eq.~\ref{eq:eta-phi} and get a set of $\phi_\mathrm{c}$ corresponding to different $\chi_\mathrm{f}$. We now fit these data to Eq.~\ref{eq:phic} with $g(\chi_\mathrm{f})=\chi_\mathrm{f}^{b_{\phi_\mathrm{c}}}$ and $b_{\phi_\mathrm{c}}\simeq2$. We apply same strategy to $k$ and $m$ and get $b_k\simeq0.4$ and $b_m\simeq2.2$ respectively (fits are presented in Supplementary Information \cite{SI}). Since we have expressions for $k(\chi_\mathrm{f})$, $\phi_\mathrm{c}(\chi_\mathrm{f})$ and $m(\chi_\mathrm{f})$, we now plot Eq.~\ref{eq:eta-chi} with corresponding $\chi_\mathrm{f}$. The plots are presented in Fig.~\ref{fig:eta_phi_frac} as dashed lines, which shows that Eq.~\ref{eq:eta-chi} fit well with simulation data. This indicates that our assumption works well to predict viscosity for given $\phi$ and $\chi_\mathrm{f}$. 

\begin{figure}
\subfigure{\includegraphics[scale=0.4]{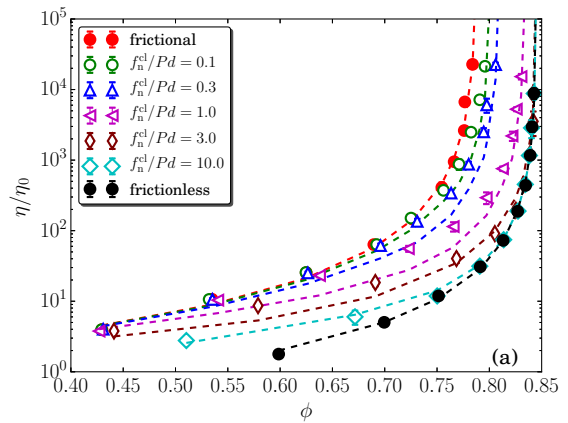}}
\subfigure{\includegraphics[scale=0.4]{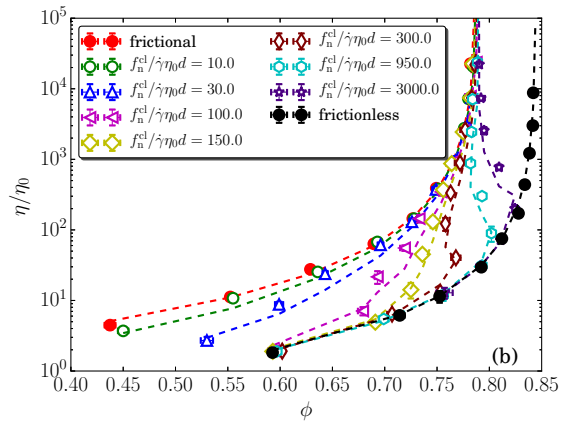}}
\caption{\label{fig:eta_phi} Viscosity $\eta/\eta_0$ as a function of packing fraction $\phi$ with (a)~varying $\dot{\gamma}, $ (b)~varying $P$; Symbols are simulation results and dashed lines are plots of Eq.~\ref{eq:eta-chi}.}
\end{figure}
Now we apply the same assumptions to the CLM model. We run simulations by varying $J$ with constant $f_\mathrm{n}^\mathrm{cl}/(Pd)$ or constant $f_\mathrm{n}^\mathrm{cl}/(\dot{\gamma}\eta_0d)$. Fig.~\ref{fig:eta_phi} shows both simulation results and plots using Eq.~\ref{eq:eta-chi} with parameters from the
binary system with $\phi$ and $\chi_\mathrm{f}$ measured from simulations with astonishing agreement and shows a robustness of these reduced parameters irrespectively of the detailed microscopic structure (see Supplementary Information \cite{SI} for two snapshots with the same $\phi$ and $\chi_\mathrm{f}$). 
For both shear protocols (keeping constant either $f_\mathrm{n}^\mathrm{cl}/(Pd)$ or $f_\mathrm{n}^\mathrm{cl}/(\dot{\gamma}\eta_0d)$), curves show a tendency of crossing over from frictional to frictionless behavior as the control parameters are changed. Both $\eta$ and $\phi$ increase monotonically for constant $f_\mathrm{n}^\mathrm{cl}/(Pd)$. Similar is found controlling  $f_\mathrm{n}^\mathrm{cl}/(\dot{\gamma}\eta_0d)$ if the parameter is low enough (typically $\lesssim 300$). On the other hand, for high and constant $f_\mathrm{n}^\mathrm{cl}/(\dot{\gamma}\eta_0d)$, one finds cusp-shape behaviors with an increasing cusp as $f_\mathrm{n}^\mathrm{cl}/(\dot{\gamma}\eta_0d)$ increases. For these cups one finds two (or more) viscosities for each packing fraction (except possibly at the maximum $\phi$). These two states do however have different pressures (assuming fixed $f_\mathrm{n}^\mathrm{cl}$), with a higher pressure for the high viscosity case. This indicates a range of pressures where one has a negative \textit{dynamic} compressibility, \textit{i.e.} $\phi$ decreases
with increasing $P$.  This appears when $\phi$ is above $\phi_\mathrm{c}^\mathrm{f}$ since then $\phi_\mathrm{c}^\mathrm{f}<\phi < \phi_\mathrm{c}\leq \phi_\mathrm{c}^\mathrm{nf}$; as $\chi_\mathrm{f}$ increases with $P$ (see Fig. 2b) so must $\phi_\mathrm{c}\to \phi_\mathrm{c}^\mathrm{f}$, yielding a negative \textit{dynamic} compressibility.

\begin{figure}
\includegraphics[scale=0.4]{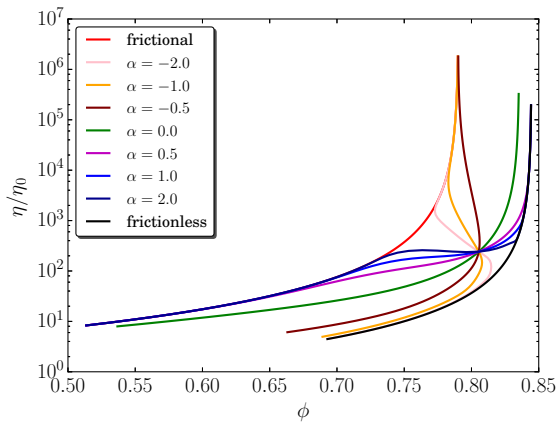}
\caption{\label{fig:eta_phi_fit} Viscosity $\eta/\eta_0$ as a function of packing fraction $\phi$ calculated from Eq.~\ref{eq:eta-J} for various $\alpha$ with $J_0=0.001$, $D=1$; frictional and frictionless curves are produced by setting $D=0$ and $\infty$.}
\end{figure}

Unlike in our reference binary system, $\chi_\mathrm{f}$ can no longer be considered as control parameter in the CLM model.
Instead it will vary with $J$ and the control parameter as determined by the shear protocol. It is reasonable to presume
that $\chi_\mathrm{f}$ is related to  $f_\mathrm{n}^\mathrm{cl}/\langle f_\mathrm{n} \rangle $ according to the presumption of the CLM model, where $\langle f_\mathrm{n} \rangle$ is the average of normal force between two disks.
Given that $P \sim \langle f_\mathrm{n} \rangle$, we assume $\chi_\mathrm{f}=\chi_\mathrm{f}(f_\mathrm{n}^\mathrm{cl}/(Pd), J)$. As $\lim_{J\to0} \phi=\phi_\mathrm{c}$ and given that $\phi_\mathrm{c}=\phi_\mathrm{c}(\chi_\mathrm{f})$, we can assume $\phi=\phi(f_\mathrm{n}^\mathrm{cl}/(Pd), J)$. We obtain expression for $\chi_\mathrm{f}$ and $\phi$ by fitting corresponding simulation data (in Supplementary Information \cite{SI}). Now we can rewrite Eq.~\ref{eq:eta-chi} as

\begin{equation}\label{eq:eta-J}
\eta=\eta(f_n^{cl}/Pd, J).
\end{equation}

Remembering that $J=\dot{\gamma}\eta_0/P$ one can reformulate both control parameters in functions of $J$ and $f_\mathrm{n}^\mathrm{cl}/(Pd)$:  $f_\mathrm{n}^\mathrm{cl}/(Pd)=(f_\mathrm{n}^\mathrm{cl}/(Pd)){J^0}$ and $f_\mathrm{n}^\mathrm{cl}/(\dot{\gamma}\eta_0d)=(f_\mathrm{n}^\mathrm{cl}/(Pd)){J^{-1}}$.
This shows the generality of Eq.~\ref{eq:eta-J}.
Following the above reasoning we actually find a whole family of shear protocols that can be encoded in the new parameter:
 \begin{equation}\label{eq:D}
 D=\frac{f_\mathrm{n}^\mathrm{cl}}{Pd}\Big(\frac{J}{J_0}\Big)^{\alpha},
 \end{equation}
where $J_0$ is a reference point. With a given $\alpha$, we can obtain relation between $J$ and $P$ which can then be put into Eq.~\ref{eq:eta-J}. Different $\alpha$'s can be divided into three parts around the two previous reference points $-1$ and $0$, which correspond to varying $\dot{\gamma}$ or varying $P$. The other $\alpha$'s indicate that both $\dot{\gamma}$ and $P$ are varied simultaneously. From Eq.~\ref{eq:D} we see that pressure should be varied as $P \sim \dot{\gamma}^{\alpha/(1+{\alpha})}$ and noticing that $J \sim \dot{\gamma}/P$ we can obtain $J \sim \dot{\gamma}^{1/(1+\alpha)}$. To observe divergence in viscosity (\textit{i.e.} $J\to0$ and consequently, $\phi\to\phi_c$),  there are three possibilities to realize this. In the first case, both $\dot{\gamma}$ and $P$ increases with $P$ increasing more rapidly, which corresponds to $\alpha < -1$. In the second case, $\dot{\gamma}$ decreases while $P$ increases, which corresponds to $-1 < \alpha < 0$. In the third cases, both $\dot{\gamma}$ and $P$ decreases with $P$ decreasing faster, which corresponds to $\alpha > 0$.
These combination of $\dot{\gamma}$ and $P$ can be easily mapped to different shear protocol in experiments. Fig.~\ref{fig:eta_phi_fit} illustrates plots of Eq.~\ref{eq:eta-J} with $J_0=0.001$ and $D=1$ for various $\alpha$'s. Frictional and frictionless curves are also produced by setting $D=0$ and $\infty$ respectively. The intersection point is controlled by both $J_0$ and $D$. Negative $\alpha$ causes $\phi$ to behave non-monotonically while positive $\alpha$ causes $\eta$ to become non-monotonic. Cusped curves become more significant with larger $|\alpha|$. 

In this work, we simulate shear thickening of dense suspensions under confining pressure. By plotting $\eta(\phi)$ and fitting the data to the empirical equation $\eta/\eta_0=\eta(\phi, \chi_\mathrm{f})$, we illustrate that both $\chi_\mathrm{f}$ and $\phi$ determine the viscosity.
In particular, the jamming point $\phi_\mathrm{c}$ is determined by $\chi_\mathrm{f}$ and hence the divergence of viscosity. We find different transition paths between the fully frictional and frictionless curves depending on the shear protocol. Negative \textit{dynamic} compressibility is observed when varying $P$ at $\phi>\phi_\mathrm{c}^\mathrm{f}$, resulting from a decreasing $\phi_\mathrm{c}$ due to increasing $\chi_\mathrm{f}$. We further find that both $\chi_\mathrm{f}$ and $\phi$ are functions of $f_\mathrm{n}^\mathrm{cl}/(Pd)$ which indicates that a relation between repulsive forces and confining pressure is a key factor of shear thickening behavior and could be used in continuum modelling. We also introduce a new dimensionless constant $D=\frac{f_\mathrm{n}^\mathrm{cl}}{Pd}(\frac{J}{J_0})^{\alpha}$. Encoded in $\alpha$ is a whole set of different shear protocols, each which generating its own flow curve. This observation could be tested and verified using current experiment protocols for shear-thickening suspensions \cite{Boyer1}. 
\newline
\acknowledgments
M.T. acknowledges financial support by the Swedish Research Council (621-2014-4387) and Crafoord Foundation (20160568). The simulations were performed on resources provided by the  Swedish National Infrastructure for Computing  (SNIC) at the center for scientific and technical computing at Lund University (LUNARC).  J.D. acknowledges Jasenko Gavran for initial help and instruction running the binary systems.

\bibliography{myref}

\clearpage

\begin{appendices}

\begin{figure}[h]
\includegraphics[scale=0.4]{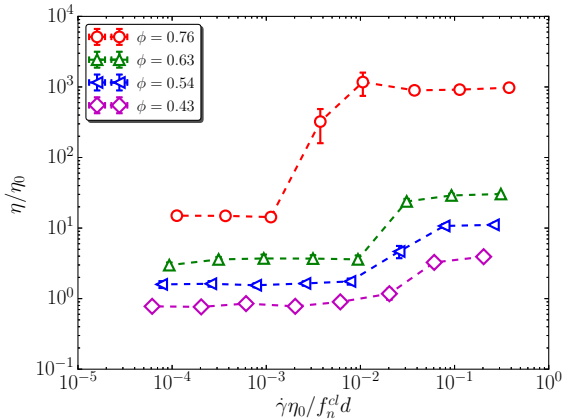}
\caption{\label{fig:eta_gamma_V} Viscosity $\eta$ as a function of  shear rate $\dot{\gamma}$ using normalised units in a log-log scale at different constant $\phi$'s. }
\end{figure}

\begin{figure*}
\subfigure{\includegraphics[scale=0.4]{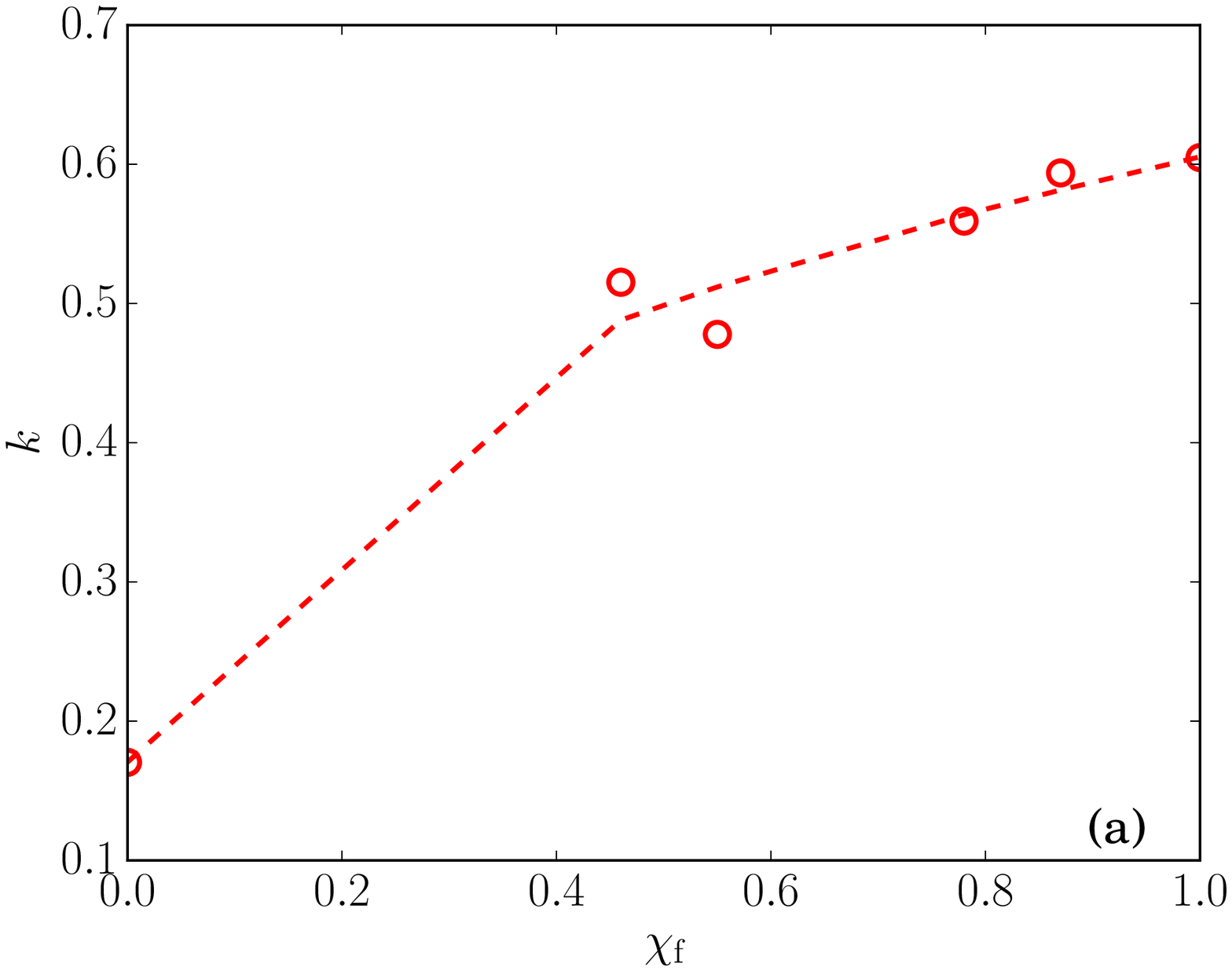}}
\subfigure{\includegraphics[scale=0.4]{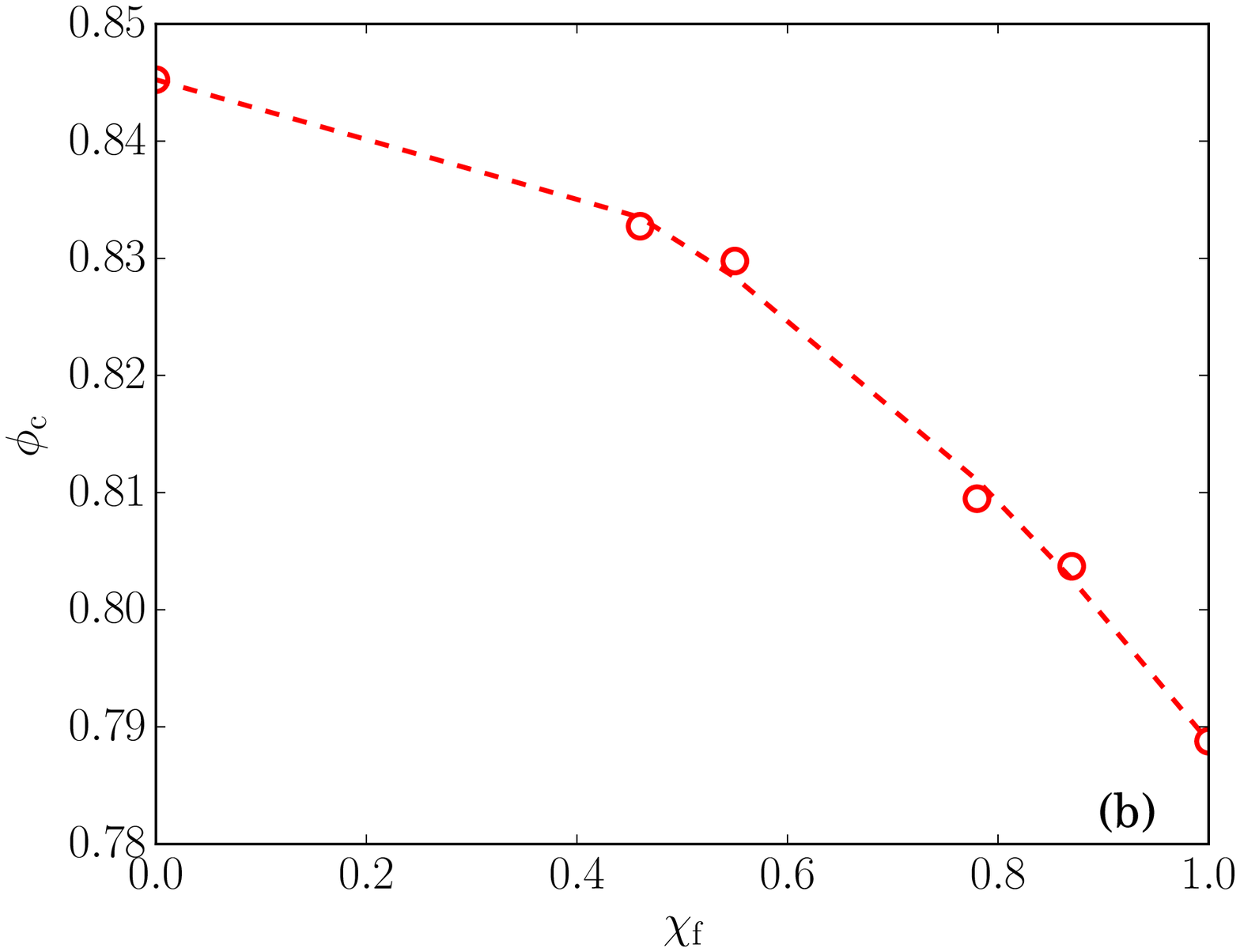}}
\subfigure{\includegraphics[scale=0.4]{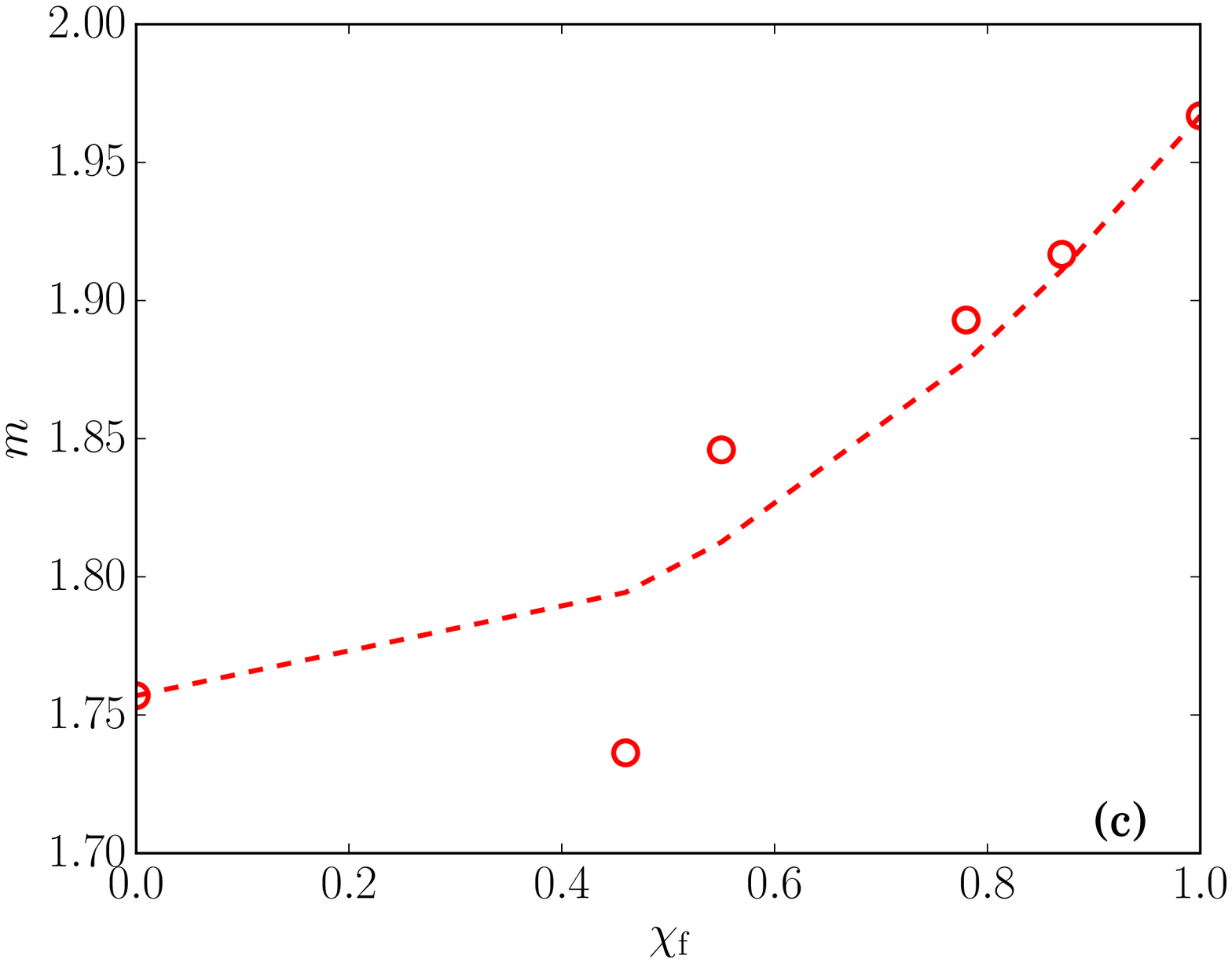}}
\caption{\label{fig:coeff-frac} Dependence of (a)~$k$, (b)~$\phi_\mathrm{c}$, and (c)~$m$ in Eq.~\ref{eq:eta-phi} on $\chi_\mathrm{f}$; the values of the parameters are obtained from the best fits of simulation data; symbols are values of the parameters and dashed lines are plots of fits.}
\end{figure*}

\begin{figure*}
\subfigure{\includegraphics[scale=0.4]{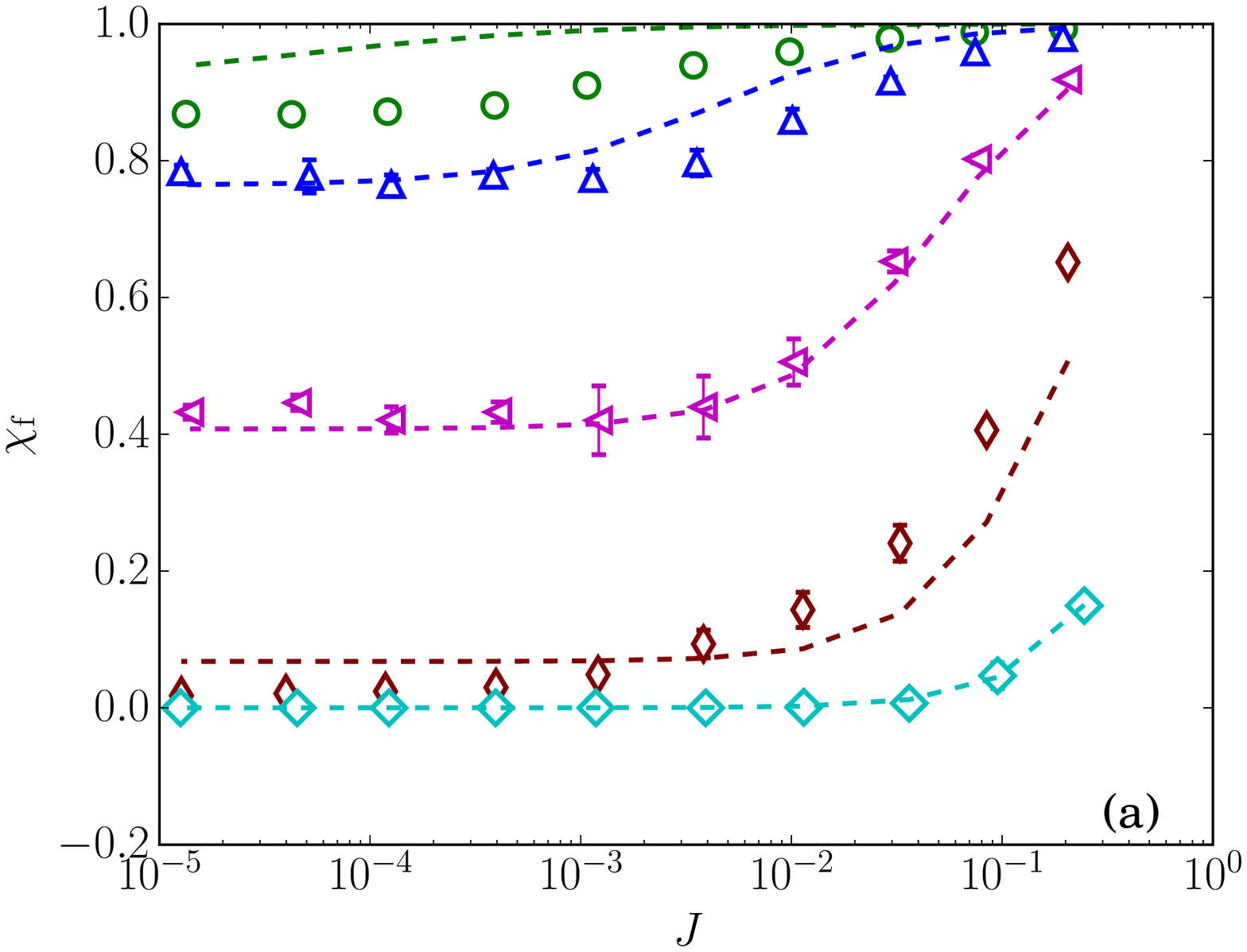}}
\subfigure{\includegraphics[scale=0.4]{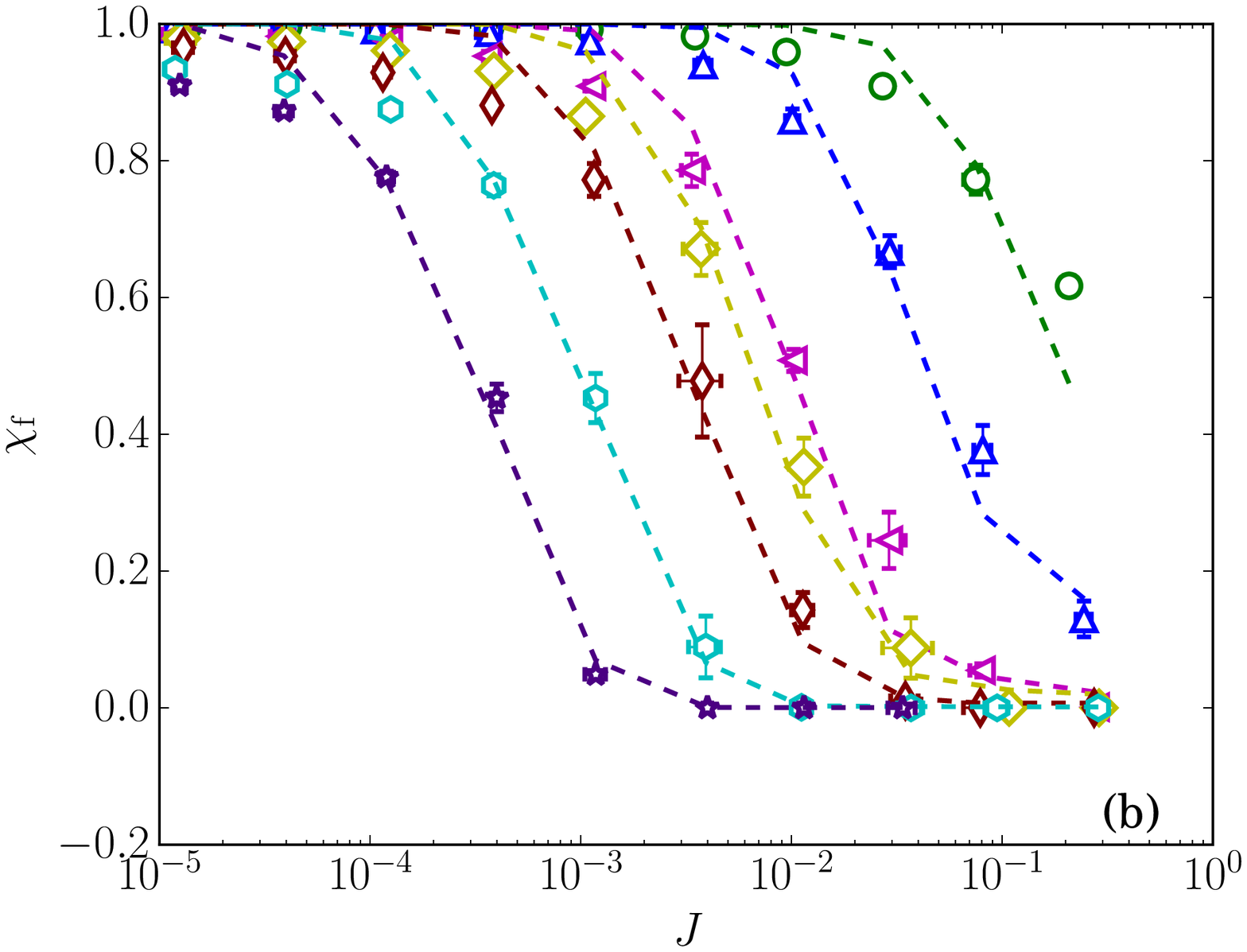}}
\caption{\label{fig:fric-J} Fraction of frictional contact $\chi_\mathrm{f}$ as a function of viscous number $J$ for various $f_\mathrm{n}^\mathrm{cl}$ with (a)~varying $\dot{\gamma}$, (b)~varying $P$; colours and styles of symbols are consistent with Fig.~\ref{fig:eta_phi}; symbols are simulation data and dashed lines are plots of Eq.~\ref{eq:chi_f}. }
\end{figure*}

\begin{figure*}
\subfigure{\includegraphics[scale=0.4]{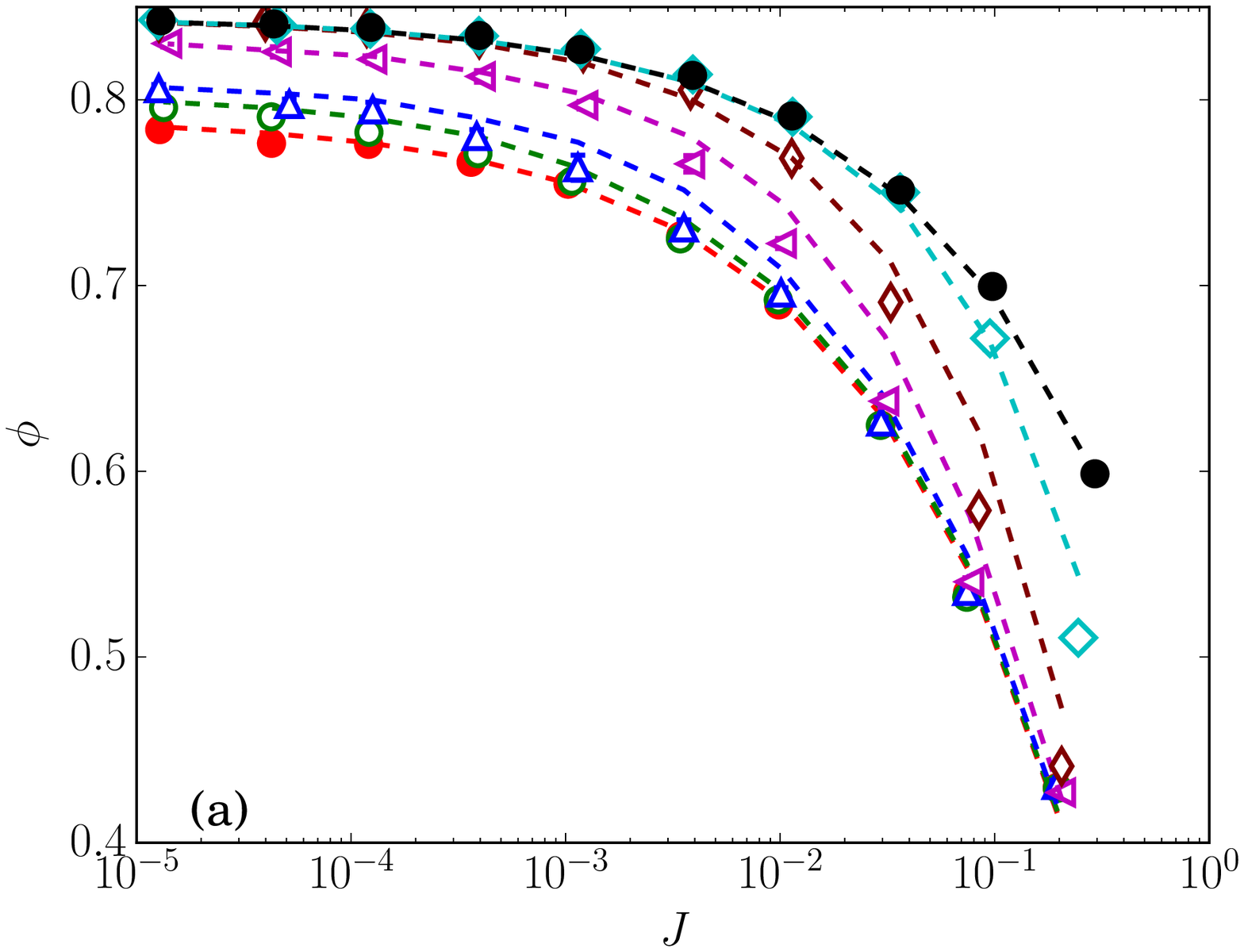}}
\subfigure{\includegraphics[scale=0.4]{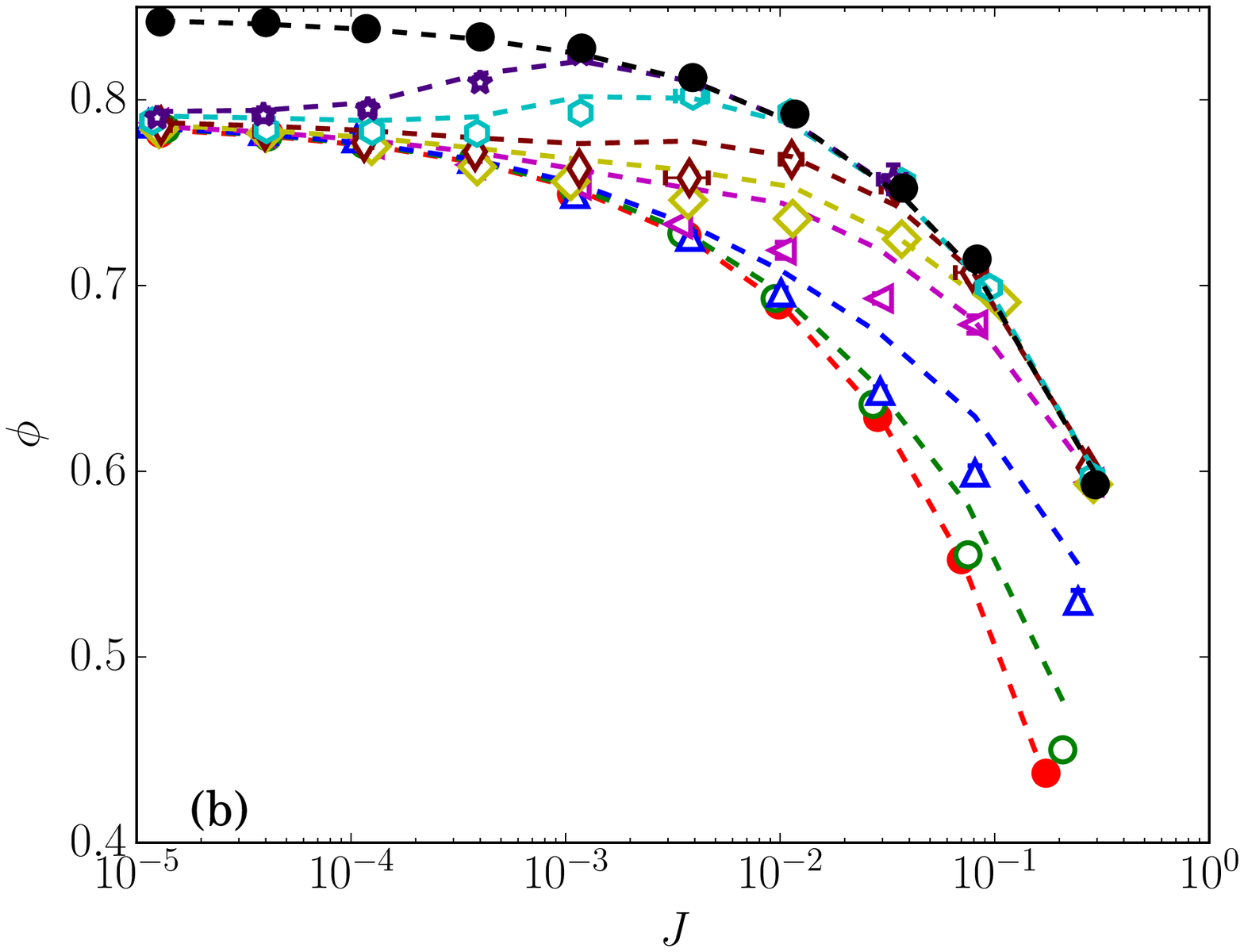}}
\caption{\label{fig:phi-J} Packing fraction $\phi$ as a function of viscous number $J$ for various $f_\mathrm{n}^\mathrm{cl}$ with (a)~varying $\dot{\gamma}$, (b)~varying $P$; colours and styles of symbols are consistent with Fig.~\ref{fig:eta_phi}; symbols are simulation data and dashed lines are plots of Eq.~\ref{eq:phi}.}
\end{figure*}

\begin{figure*}
\subfigure{\includegraphics[scale=0.4]{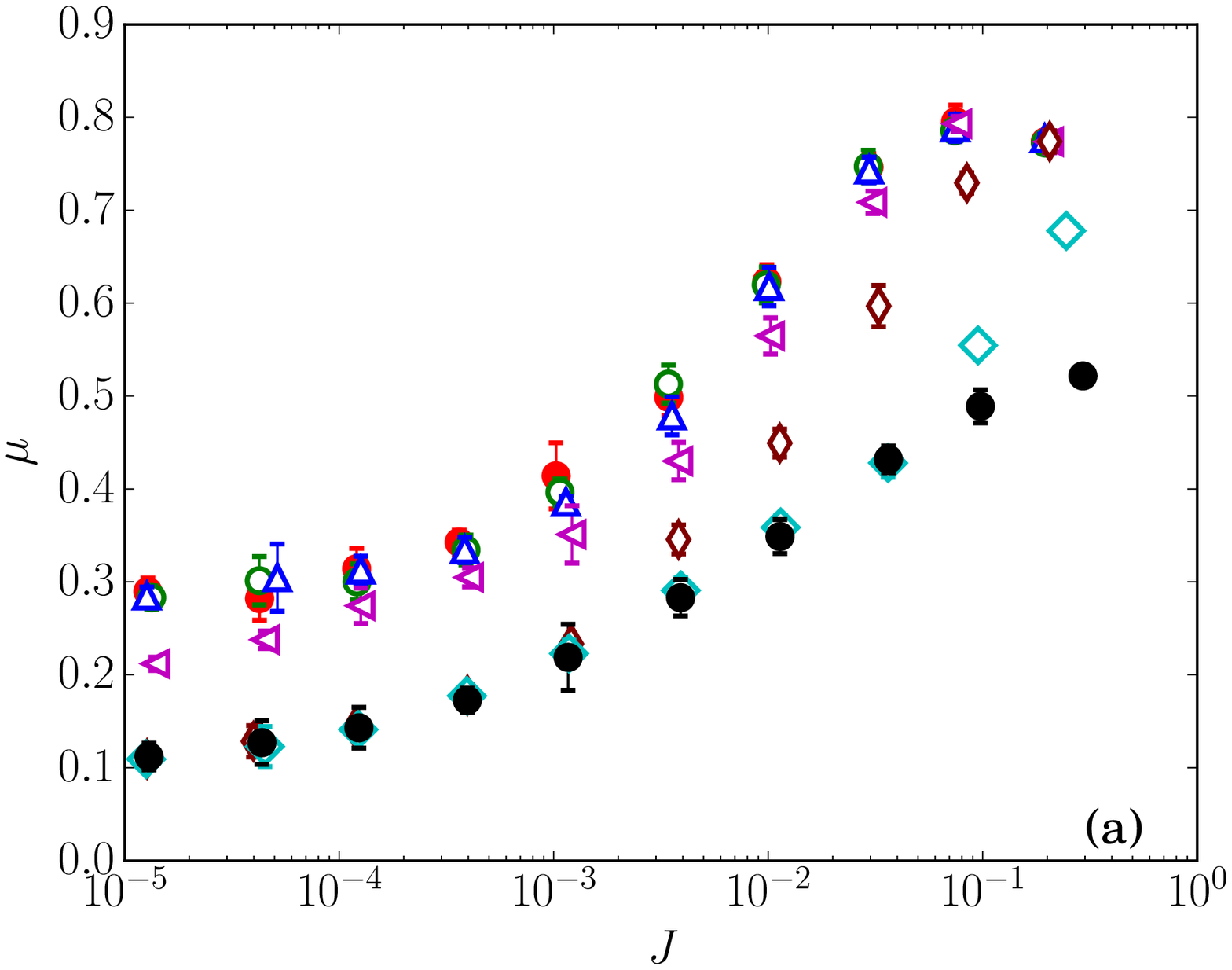}}
\subfigure{\includegraphics[scale=0.4]{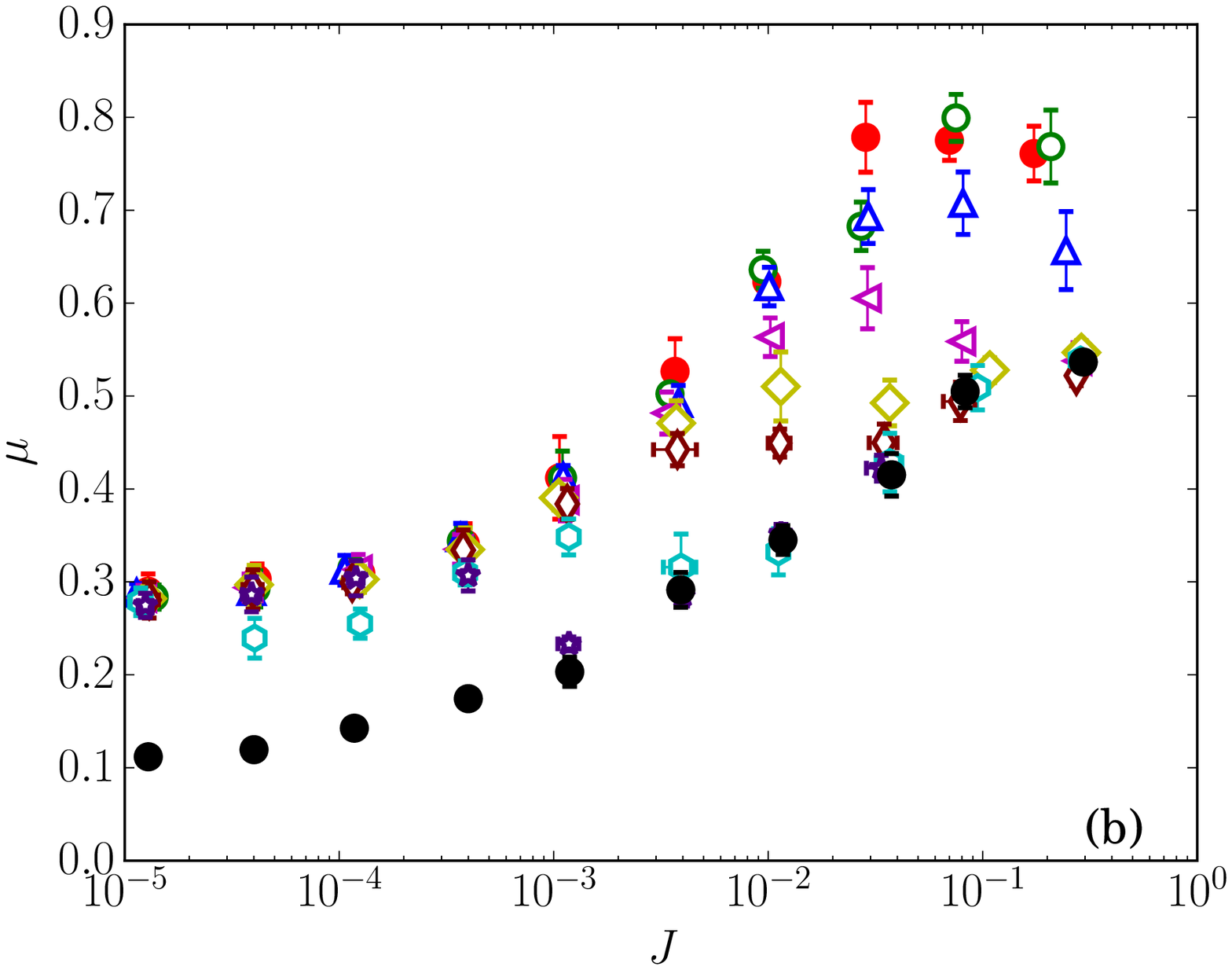}}
\caption{\label{fig:mu-J} Macroscopic friction $\mu$ as a function of $J$ for various $f_\mathrm{n}^\mathrm{cl}$ with (a)~varying $\dot{\gamma}$, (b)~varying $P$; colours and styles of symbols are consistent with Fig.~\ref{fig:eta_phi}.}
\end{figure*}

Fig.~\ref{fig:eta_gamma_V} illustrates the dependence between viscosity $\eta$ and shear rate $\dot{\gamma}$ at various constant $\phi$'s. 
Fig.~\ref{fig:coeff-frac} shows curves for $k(\chi_\mathrm{f})$, $\phi_\mathrm{c}(\chi_\mathrm{f})$, and $m(\chi_\mathrm{f})$. The fitting function has been described before. Fig.~\ref{fig:fric-J} shows how $\chi_\mathrm{f}$ varies with increasing $J$ using different shear protocols. The data in Fig.~\ref{fig:fric-J}(a) are fit into 
\begin{equation}\label{eq:chi_f}
\chi_f=k_1+(1-k_1)\cdot\frac{J^{k_2}}{k_3+J^{k_2}},
\end{equation}
where $k_1$, $k_2$ and $k_3$ are all functions of $f_\mathrm{n}^\mathrm{cl}/P$. After getting expressions for $k_1$, $k_2$ and $k_3$, we obtain expression for $\chi_\mathrm{f}=\chi_\mathrm{f}(f_\mathrm{n}^\mathrm{cl}/P, J)$. Plots of $\chi_\mathrm{f}=\chi_\mathrm{f}(f_\mathrm{n}^\mathrm{cl}/P, J)$ for both protocols are illustrated in Fig.~\ref{fig:fric-J} as dashed lines. The expression for $\phi=\phi(\chi_\mathrm{f}, J)$ is obtained with same strategy. The fitting function is 
\begin{equation}\label{eq:phi}
\phi=\phi_\mathrm{c}+k'J^{0.45},
\end{equation}
where $k'$ is a function of $\chi_\mathrm{f}$.  The results are shown in Fig.~\ref{fig:phi-J} where dashed lines are plots of Eq.~\ref{eq:phi} and symbols are simulation data. 
Fig.~\ref{fig:mu-J} shows how macroscopic friction $\mu$ varies with $J$ using two shear protocols. 

\begin{figure*}
\subfigure{\includegraphics[scale=0.4, trim={2cm 6cm 2cm 6cm},clip]{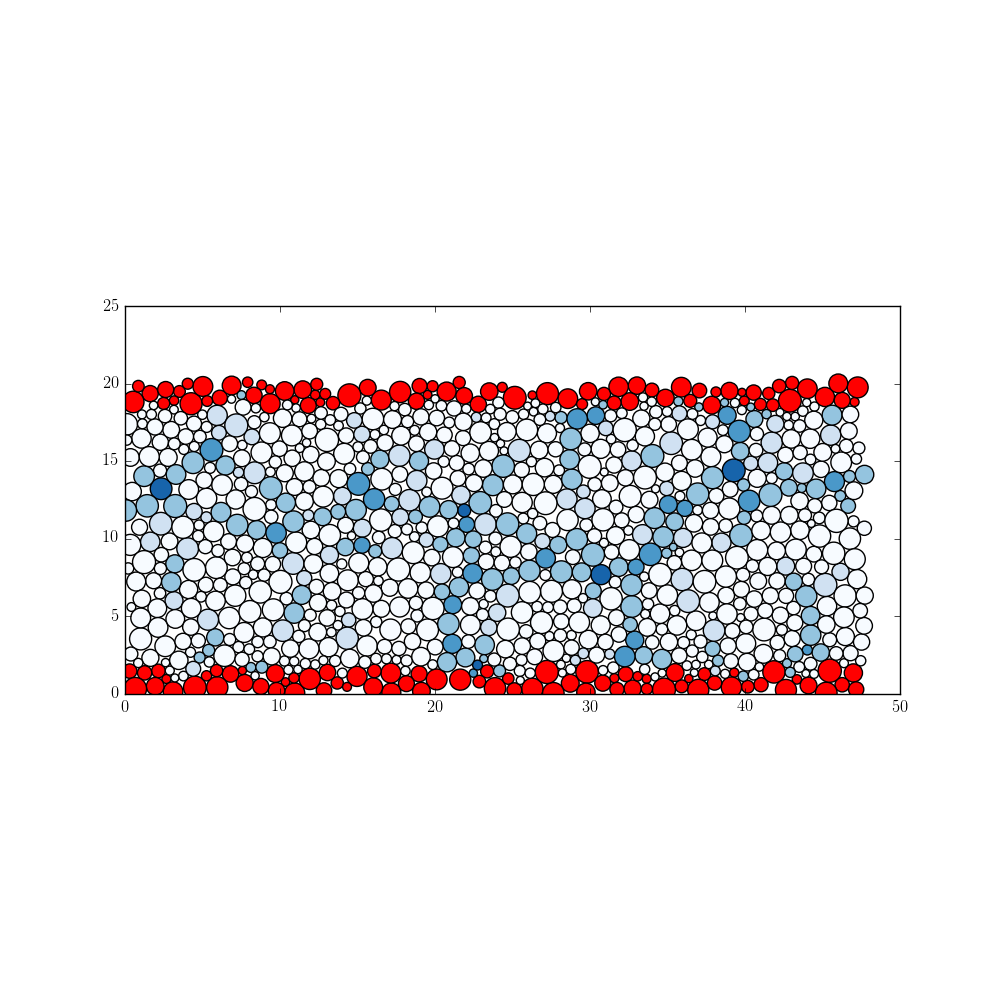}}
\subfigure{\includegraphics[scale=0.4, trim={2cm 6cm 2cm 6cm},clip]{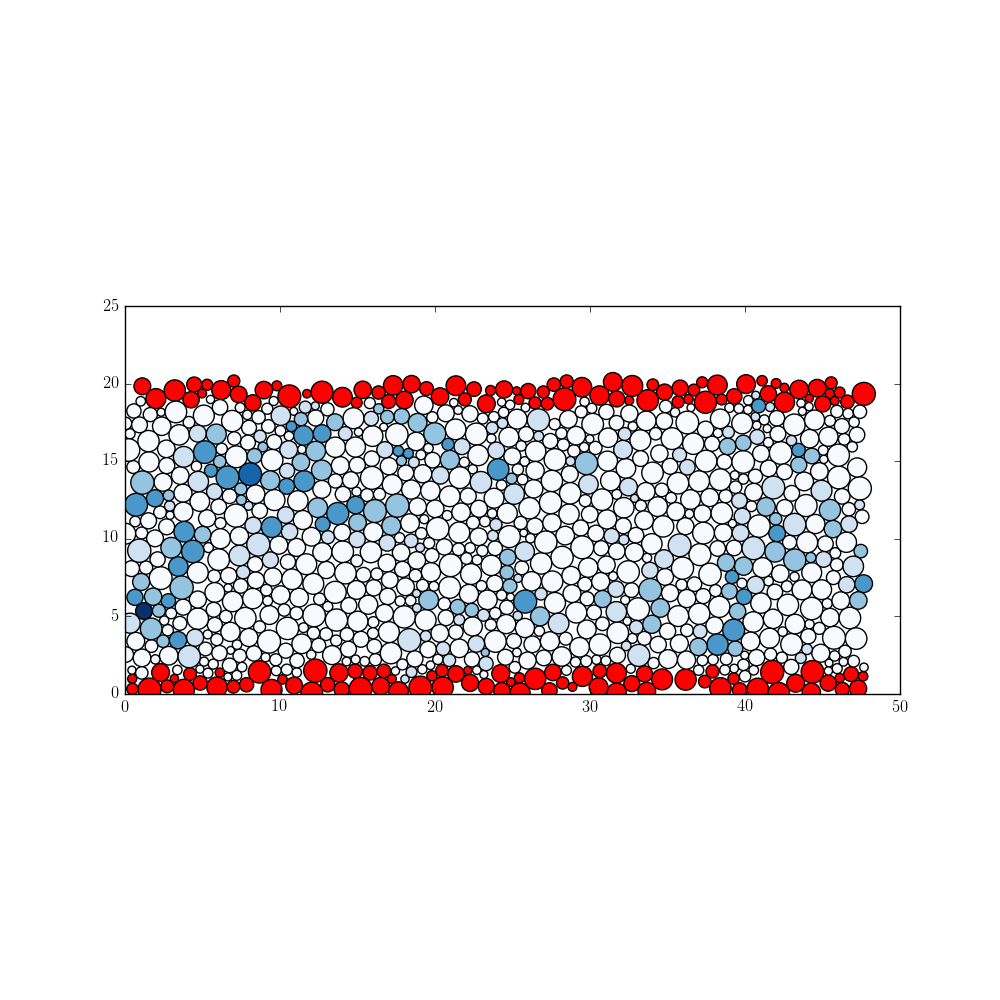}}
\caption{\label{fig:snapshots} Snapshots for (left)~CLM model, (right)~binary system; red circles are wall disks, white circles are disks with no frictional contacts, and blue circles represents disks with frictional contact and darker colour indicates larger number of frictional contacts.}
\end{figure*}

Fig.~\ref{fig:snapshots} shows microscopic structure of suspensions under shearing using CLM model and binary system (two animations are also attached). Corresponding $\chi_\mathrm{f}$ and $\eta$ are $\chi_\mathrm{f}^\mathrm{CLM}\simeq0.12$, $\chi_\mathrm{f}^\mathrm{binary}\simeq0.13$; $\eta^\mathrm{CLM}\simeq9500$, $\eta^\mathrm{binary}\simeq10500$. Although $\chi_\mathrm{f}$ and $\eta$ for two different systems are quite close to each other, the microscopic structures are rather different as illustrated in Fig.~\ref{fig:snapshots}. For system using CLM model, disks with frictional contacts form chains or even networks while for binary system, these disks are distributed more separately.  This observation shoes the subdominant effect of the chains for viscosity.

\end{appendices}

\end{document}